\documentclass[fleqn,usenatbib]{mnras}

\usepackage{mathptmx}

\usepackage[T1]{fontenc}

\DeclareRobustCommand{\VAN}[3]{#2}
\let\VANthebibliography\thebibliography
\def\thebibliography{\DeclareRobustCommand{\VAN}[3]{##3}\VANthebibliography}

\usepackage{graphicx}	
\usepackage{amsmath}	
\usepackage{amssymb}	
\usepackage{natbib}	

\title[The updated BaSTI white dwarf models]{The updated BaSTI stellar evolution models and isochrones. III. White Dwarfs}

\author[M.Salaris et al.]{
Maurizio Salaris,$^{1,2}$\thanks{E-mail: M.Salaris@ljmu.ac.uk}
Santi Cassisi,$^{2,3}$
Adriano Pietrinferni$^{2}$
and Sebastian Hidalgo$^{4,5}$
\\
$^{1}$Astrophysics Research Institute, Liverpool John Moores University, 146 Brownlow Hill, Liverpool, L3 5RF, UK\\
$^{2}$INAF -- Osservatorio Astronomico di Abruzzo, Via M. Maggini, s/n, I-64100, Teramo, Italy\\
$^{3}$INFN -- Sezione di Pisa, Largo Pontecorvo 3, 56127 Pisa, Italy\\
$^{4}$Instituto de Astrofisica de Canarias, Via Lactea s/n, La Laguna, Tenerife, Spain\\
$^{5}$Department of Astrophysics, University of La Laguna, Via Lactea s/n, La Laguna, Tenerife, Spain
}

\date{Accepted XXX. Received YYY; in original form ZZZ}

\pubyear{2015}

\begin{document}
\label{firstpage}
\pagerange{\pageref{firstpage}--\pageref{lastpage}}
\maketitle

\begin{abstract}
  We present new cooling models for carbon-oxygen white dwarfs with both H- and He-atmospheres, covering the whole relevant
  mass range, to extend our updated BaSTI (a Bag of Stellar Tracks and Isochrones)
  stellar evolution archive. They have been computed using  
  core chemical stratifications obtained from new progenitor calculations, adopting a semiempirical initial-final mass relation.
  The physics inputs have been updated compared to our previous BaSTI calculations: $^{22}$Ne diffusion
  in the core is now included, together with an updated CO phase diagram,
  and updated electron conduction opacities. We have calculated models with various different neon abundances in the core, 
  suitable to study white dwarfs in populations with 
  metallicities ranging from super-solar to metal poor, and have performed various tests/comparisons of the chemical stratification
  and cooling times of our models. 
  Two complete sets of calculations are provided, for two different choices of the electron conduction opacities,
  to reflect the current uncertainty in the evaluation of the electron thermal conductivity in the
  transition regime between moderate and strong degeneracy, crucial for the H and He envelopes.
  We have also made a first, preliminary estimate of the effect --that turns out to be generally small --
  of Fe sedimentation on the cooling times of white dwarf models, following recent calculations of
  the phase diagrams of carbon-oxygen-iron mixtures.
  We make publicly available the evolutionary tracks from both sets of calculations, including cooling times and magnitudes in
  the Johnson-Cousins, Sloan, Pan-STARSS, Galex, $Gaia$-DR2, $Gaia$-eDR3, $HST$-ACS, $HST$-WFC3, and $JWST$
  photometric systems.
 \end{abstract}




\section{Introduction}

According to theory,  
white dwarfs (WDs) are the most common final stage of the evolution of single stars. Indeed, all single stars 
with initial mass up to $\sim$9-10$M_{\odot}$ are predicted to end up as either He-core (for initial masses below
$\sim$0.5$M_{\odot}$), CO-core (initial masses between $\sim$0.5 and $\sim$6-7$M_{\odot}$), or ONe-core 
(initial masses between 6-7 and 9-10$M_{\odot}$) WDs, and 
due to the current age of the universe and the shape of the stellar initial mass function,
the overwhelming majority of WDs produced so far by single star evolution are predicted to have a CO core.

In the last 25 years, thanks to steady advances in both observations and theory, 
models of CO-core WDs have been used extensively in conjuction with photometric, spectroscopic and
asteroseismic data, to determine the ages of 
field WDs \citep[e.g.,][]{winget, oswalt, torres16, kilic17, tononi} for constraining the star formation history of
the Milky Way, the ages of WDs in open clusters
\citep[e.g.,][]{m67richer,vonHippel,ngc6791,m67,garciaberro,ngc2158,ngc6819} and 
globular clusters
\citep[e.g.,][]{m4hansen,ngc6397,winget3,m4,ratecooling,ngc6752}, and even as probes 
to investigate open questions in theoretical physics 
\citep[e.g.,][]{freese, primero, gdot1, corsico, gdot3, bertone, gdot2, jordi18, winget2}.

As part of an ongoing new release of the BaSTI (a Bag of Stellar Tracks and Isochrones) stellar model library
\citep{bastiiacss, bastiiacae}, and
given the important role played by evolutionary WD models to address a range of major astrophysics and physics questions,
we present here new sets of CO-core WD model computations, that update the 
\citet{bastiwd} calculations --hereafter S10-- part of the previous BaSTI release, taking advantage of recent improvements
in a number of stellar physics inputs.
In these new calculations, that cover the whole relevant mass range,
we have included the diffusion of $^{22}$Ne in the liquid phase, which wasn't accounted for
by S10, we have updated the electron conduction opacities, the phase diagram for CO compositions, and also the CO
stratification in the cores of the models, as described in the next sections.

Following the recent analysis by \citet{cpsp}, we present calculations for
two different choices of the electron conduction opacities, and we also provide a preliminary 
estimate of the effect on the model cooling times of
iron sedimentation in the CO cores, based on the very recent study by 
\citet{caplanFe} of the phase diagram of carbon, oxygen, and iron ternary mixtures.
We make publicly available at the official BaSTI website\footnote{\url{http://basti-iac.oa-abruzzo.inaf.it}} the evolutionary tracks (cooling tracks) in the Hertzsprung-Russell diagram,
including cooling times and magnitudes in several photometric filters, obtained for all WD
masses and chemical stratifications presented here.

The plan of the paper is as follows. Section~\ref{mod} describes the parameter space covered by our WD calculations, 
the adopted physics inputs and initial chemical abundance profiles, and the effect of the $^{22}$Ne diffusion on both the
chemical stratification and cooling times of the models.
The following Sect.~\ref{test_profiles} presents numerical tests 
to quantify the effect of progenitor metallicity and the adopted relation between progenitor mass and final WD mass,
on chemical profiles and cooling times. We also discuss how the cooling times are affected
by considering the idealized, 
sharp chemical transitions between the CO core and the pure-He/H envelopes adopted in our models (and other 
models in the literature), and by neglecting the small fraction of heavy elements and 
atomic diffusion in the envelopes during the WD evolution.
Section~\ref{compopa} compares the cooling times obtained with the two choices for the conductive
opacities, while
Sect.~\ref{iron} presents a first preliminary estimate of the effect of iron sedimentation of our model cooling times.
Section~\ref{other} compares cooling times and radii of our new calculations with models from the literature, and is
followed by 
a brief summary of our results which brings the paper to a close.

\section{Models}
\label{mod}

We have calculated CO-core WD cooling models with pure-H (representative of H-atmosphere WDs) and pure
He (representative of He-atmosphere WDs) envelopes, with masses $M_{\rm WD}$ equal to 
0.54, 0.61, 0.68, 0.77, 0.87, 1.0 and 1.1 $M_{\odot}$, respectively. These masses are in common with our previous 
S10 calculations, apart from the highest value.

All computations include the important processes of diffusion of $^{22}{\rm Ne}$ in the liquid
phase of the CO-cores \citep[see][]{bravo,bh01,db02,gb08,althaus10}, latent heat release 
and phase separation upon crystallization
\citep[see, e.g.,][]{stevenson, m83, garciaberro88}, which provide sizable contributions to the
energy budget of the models.

As in our previous sets of calculations \citep[][and S10]{s00}, 
for each mass an initial model has been converged 
at $\log(L/L_{\odot}) \sim 1.0-1.5$, by considering a reference
CO-core chemical stratification and a reference thickness and
chemical composition of the envelope layers, described in Sect.~\ref{chemistry}. 

\begin{figure}
	\includegraphics[width=\columnwidth]{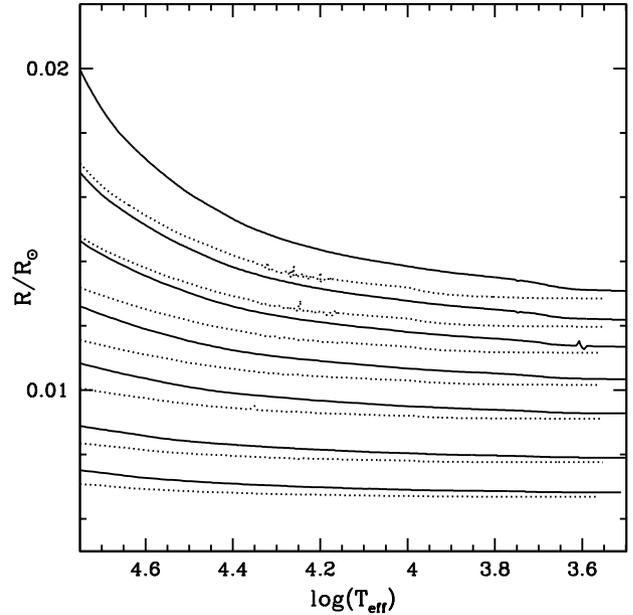}
        \caption{Radius of our grid of H-atmosphere (solid lines) and He-atmosphere (dotted lines)
          models calculated with \citet{cas07} electron conduction opacities, as a function of the models' $T_{\mathrm eff}$
        (at fixed $T_{\mathrm eff}$ larger radii correspond to less massive models).}
    \label{fig:mr}
\end{figure}

\begin{figure}
	\includegraphics[width=\columnwidth]{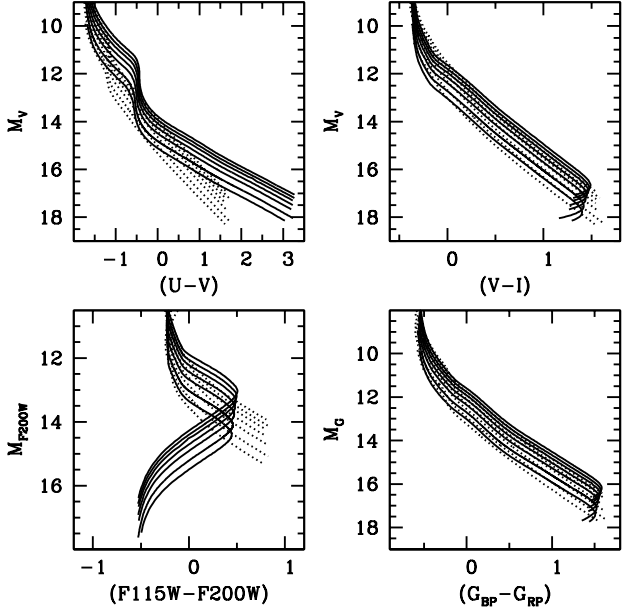}
        \caption{Selected CMDs of the models in Fig.~\ref{fig:mr} (solid lines for H-atmosphere models, dotted lines
          for He-atmosphere models) in Johnson-Cousins (upper panels), 
          $HST$-WFC3 (bottom-left panel), and $Gaia$-eDR3 (bottom-right panel) filters.
        }
    \label{fig:cmd}
\end{figure}

For each mass and core chemical stratification we have computed models for two different choices of
the electron conduction opacities --from \citet{cas07} and \citet{b20}, respectively-- as described in Sect.~\ref{physics}.
Figures~\ref{fig:mr} and ~\ref{fig:cmd} display, as an example, the radius of our grid of H-atmosphere and He-atmosphere
models calculated with \citet{cas07} opacities
as a function of the effective temperature $T_{\mathrm eff}$, and the resulting cooling tracks in 
selected colour magnitude diagrams (CMDs). We can easily notice in three of the displayed CMDs
of H-atmosphere models the turn to the blue of the colours at faint magnitudes 
(more pronounced in infrared CMDs, like the $HST$-WFC3 ($F115W,F200W$) diagram)
caused by the ${\rm H_2}$ collision induced absorption in the atmospheres \citep[see, e.g.,][]{h98, sj99}.
Mass-radius relations and CMDs for all
models calculated with \citet{b20} opacities are identical to those displayed in these figures.

\subsection{Physics inputs}
\label{physics}

The WD evolution code employed for our calculations is the same described in 
\citet{s00} and S10. We briefly summarize here the main physics inputs, and what has
changed compared to S10.

The equation of state (EOS) for the H and He layers is from   
\citet{Saea95}, supplemented in the liquid phase by calculations following
\citet{segretain94} formalism; for the CO cores we adopted \citet{segretain94} calculations
in the solid and liquid phase, and \citet{straniero} EOS for the gas phase.
We have made some test calculations employing for the H and He layers the recent EOS published by \citet{cms},
  an update of the \citet{Saea95} one. We have found 
  negligible differences in the structure and cooling times of the models, compared to computations with the 
  \citet{Saea95} EOS.

Phase separation upon crystallization of the CO mixture is included employing the
new phase diagram by \citet{bd21}, and the energy contribution due to phase separation is calculated
following \citet{isern00}.

The diffusion of $^{22}{\rm Ne}$ (hereafter Ne) in the liquid layers of the CO cores is
included following \citet{gb08} and \citet{althaus10}.
Neon is considered to diffuse as a trace element in a one-component plasma background
-- see Eqs. 19 and 20 in \citet{gb08}-- 
made of a fictitious element with mean atomic weight and charge calculated by averaging over 
the local number fractions of C and O.
The Ne diffusion coefficients come from the most recent results by \citet{hughto},
and the contribution of this process to the energy budget has been calculated following 
\citet{gb08}.
We have not included phase separation of Ne in the core in the solid phase, a process whose occurrence 
is under debate. It is found to be efficient in WD cores 
by \citet{b21} calculations, while \citet{chc} results show that Ne cannot separate for realistic values of 
the Ne abundance in WDs.

Rates of neutrino energy losses are taken from \citet{itohneutr}, with the exception of plasma neutrino
emission, taken from the more accurate \citet{haft} calculations\footnote{The more recent
  determination of plasma neutrino emission rates by \citet{kg} provides results in agreement with
\citet{haft}, in the temperature-density regime of our calculations.}.
Radiative opacities for temperatures above 10,000~K are taken from \citet{opal},
while below 10,000~K we employed \citet{sj99} results for pure-H, and \citet{alexander}
for pure-He (and the mixed H-He or C-O-He compositions in some tests described in Sect.~\ref{test_profiles})
compositions.

Electron conduction opacities are from the calculations by \citet{cas07} in one set of models,
and from \citet{b20} for the H and He envelopes \citep[keeping][opacities for the CO cores]{cas07} in a second set.
\citet{b20} have recently published new, improved calculations of electron conduction opacities for H and He compositions
for moderate degeneracy (i.e. when $\theta \sim 1$, where $\theta\equiv T/T_{\mathrm F}$, and $T_{\mathrm F}$
is the Fermi temperature) which they have bridged 
with \citet{cas07} calculations for strong degeneracy ($\theta \lesssim 0.1$). 
These new opacities are lower by a factor up to 2.5--3 near the boundary
of the temperature-density domain where the new calculations are valid.
The treatment of the conductive opacities at the transition from moderate to 
strong degeneracy is however still uncertain --
as discussed in detail by \citet{cpsp}, see also \citet{b20}--  and is crucial for the envelopes of WD models.
Different ways to bridge \citet{b20} calculations for moderate degeneracy with \citet{cas07} for strong degeneracy
can potentially \lq{recover\rq} \citet{cas07} opacities in the regime of H and He WD envelopes, and for this reason
we have calculated models using both sets of opacities for H and He compositions.
We will show in Sect.~\ref{compopa} the main differences between these two sets of models.

Photospheric boundary conditions for $T_{\mathrm eff}$ below 10,000~K have been obtained from calculations
of the non-grey model atmospheres described
in S10, while at higher $T_{\mathrm eff}$ we use the integration of
the Eddington grey $T(\tau)$ relation.
Superadiabatic convection in the envelope is treated according to the \citet{bv} formalism of the mixing length theory, 
with mixing length $\alpha_{\mathrm MLT}$=1.5 \citep[see][]{ljs}.
Hydrogen burning through the $p-p$ cycle in the model envelopes is included, using the same reaction rates
as in \citet{bastiiacss}. As discussed later, its effect on our models is negligible.

The bolometric corrections necessary to place our WD cooling tracks in CMDs are based on the
results by \citet{bwb} and \citet{hb06}, the same employed by 
S10 \footnote{\url{https://www.astro.umontreal.ca/~bergeron/CoolingModels/}}.
For each track we have calculated magnitudes in the Johnson-Cousins, Sloan, Pan-STARSS, Galex, $Gaia$-DR2, $Gaia$-eDR3, 
$HST$-ACS, $HST$-WFC3, $JWST$ photometric systems.

To summarize, the main differences with respect to S10 regarding the physics inputs of the models are
the inclusion of Ne diffusion, the 
use of the \citet{bd21} CO phase diagram instead of \citet{sc93}, and the
\citet{cas07} \citep[and alternatively][]{b20} conductive opacities, 
instead of a combination of the older \citet{hl69} tables, and \citet{itoh1}, \citet{itoh3} analytic formulae
employed by S10.

\subsection{Chemical stratification}
\label{chemistry}

To determine the initial chemical stratification of WD models of a given mass, we need in principle to
compute the full evolution of the progenitor, from the main sequence through the challenging
thermally-pulsing asymptotic giant branch, and post asymptotic giant branch phases
\citep[see, e.g.][]{renedo,camisassa16,camisassa17}.
It is however fair to say that the 
modelling of these advanced evolutionary stages of low- and intermediate-mass stars is subject to 
non-negligible uncertainties, mainly related to the efficiency of mass loss, and internal mixing during
the thermal pulses.
For example, independent evolutionary calculations show appreciable differences in the relation between the
initial mass of the progenitor and the final WD mass 
\citep[see, e.g.,][]{mgir, wf, cummings}; given
that the CO stratification in the WD progeny depends on the initial mass of their progenitors, 
variations of the initial-final mass relation (IFMR) imply variations of the predicted CO profile for a fixed $M_{\rm WD}$.
Also, the mass thickness of the H- and He-rich  layers sorrounding the CO core depend on how many thermal pulses have been
experienced by the progenitor, the timing of the last pulse, and the efficiency of mass
loss during the post asymptotic giant branch phase
\citep[see, e.g.,][]{dge, a05, a15}, that are still difficult to predict accurately by means of stellar evolution
calculations.

Following the philosophy of S10, we provide here calculations 
with a fixed constant initial thickness of the H and He envelope layers 
for all WD masses and all initial compositions
of the progenitors \citep[see also, e.g.][for a similar approach]{hansenmodels, fbb, bedard}, and a fixed,
semiempirical IFMR to determine the core CO stratifications. 
This choice provides a reference baseline to eventually help 
disentangle the effect of the existing uncertainties
related to core and envelope physical and chemical properties, when comparisons with observations are made.

The H-atmosphere models have \lq{thick\rq} H layers, as in S10; 
the envelope is made of pure hydrogen layers enclosing a fraction $q({\rm H})$=$10^{-4}$ of the total mass $M_{\rm WD}$, 
around pure-He layers of mass $q({\rm He})=10^{-2}$, which surround the CO core
\citep[the same values chosen by][]{hansenmodels, fbb, bedard}.
The He-atmosphere WD models
have a pure helium envelope with mass fraction $q({\rm He})=10^{-2}$ around the CO core.
The chosen orders of magnitude of $q({\rm H})$ and $q({\rm He})$ are informed 
by results of evolutionary calculations of WD progenitors
\citep[see, e.g,][]{imcd, renedo}.

With these choices of $q({\rm H})$ and $q({\rm He})$ the surface convection, which eventually develops
during the cooling evolution, is not able cross the H-He interface in the H-atmosphere models,
or the He-CO interface in the He-atmosphere models. Also, $p-p$ burning slowly changes
with time the chemical composition of the deeper parts of the H envelopes of the models, producing a small (in mass)
tail of increasing He abundance when moving inwards. However, the cooling times of the models (and the total
mass of H and He in the envelope) are not changed 
appreciably compared to the case of no burning allowed.


Whilst the H and He stratification has been set at the start
of the cooling sequence without modelling the previous evolution, the profiles of the C, O and Ne abundances
in the CO cores have been taken from the evolution of models of the chosen progenitors.  
To this purpose, we have considered the semiempirical IFMR by
\citet{cummings}, and for each value of $M_{\rm WD}$ in our model grid we have derived the corresponding
value of the progenitor mass $M_{\rm i}$
from \citet{cummings} formula\footnote{We have considered \citet{cummings}
  results obtained using the \citet{parsec} models for the WD progenitor lifetimes, instead of the
  \citet{mist} ones. The differences in the predicted progenitor masses
  are however negligible for $M_{\rm WD}$ lower than 1$M_{\odot}$, and reach at most 0.6$M_{\odot}$
  for $M_{\rm WD}$=1.1$M_{\odot}$.}.

For any given $M_{\rm WD}$ we have then computed the evolution of the progenitor 
with mass $M_{\rm i}$ starting from the pre-main sequence, 
employing the same code and physics inputs of 
\citet{bastiiacss} calculations, including overshooting from the convective cores during the main sequence (but not
from the convective envelopes, when present), semiconvection
during the core He-burning phase, 
mass loss using \citet{reimers} formula with the free parameter $\eta$ set to 0.3, 
and an initial solar scaled chemical composition with metal mass fraction $Z$=0.017 
\citep[about solar --see][for details of the code]{bastiiacss}.
We are not interested in following the details of the
asymptotic giant branch evolution of these models, just in the growth of the He-exhausted
core; the initial CO stratification of the WD model has been then taken as the CO abundance profile inside
the He-free core of the progenitor, when this has reached a mass equal to $M_{\rm WD}$.

\begin{figure}
	\includegraphics[width=\columnwidth]{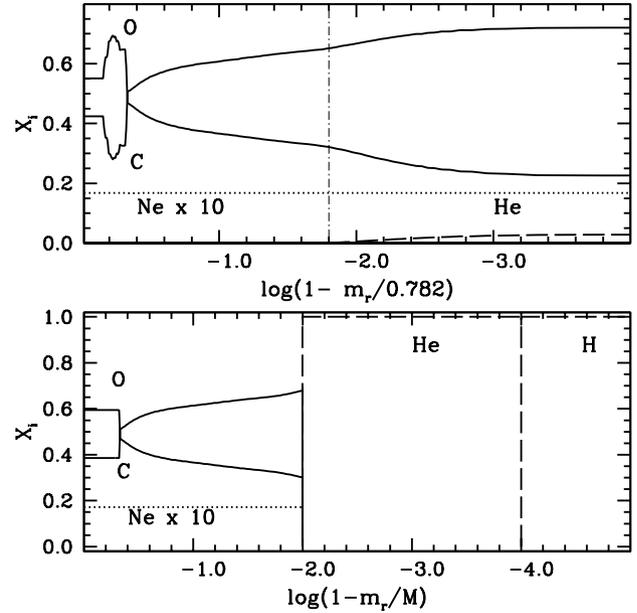}
        \caption{{\sl Upper panel: } Stratification of the  C, O, Ne, H and He abundances
          in the inner layers of a 3.1$M_{\odot}$ asymptotic giant branch model with initial chemical composition
          about solar.
          Abundances are given in mass fractions (for the sake of clarity the Ne mass fraction
          is increased by a factor of 10) and $m_r$ is the mass enclosed
          within a radial distance $r$ from the centre of the model. The stratification is taken
          around the time the He-free core (whose boundary is marked by the thin dash-dotted vertical line)
          has reached 0.77$M_{\odot}$, the final WD mass prescribed
          by the \citet{cummings} IFMR (see text for details). 
          {\sl Lower panel: } Initial chemical stratification adopted for the corresponding
          0.77$M_{\odot}$ H-atmosphere WD model ($M$ is the total mass of the model).}
    \label{fig:bastiprofile}
\end{figure}

As an example, the upper panel of Fig.~\ref{fig:bastiprofile} shows the case of the models with $M_{\rm WD}$=0.77$M_{\odot}$;
it displays the inner chemical stratification of the progenitor with mass $M_{\rm i}=3.1 M_{\odot}$  
around the time its He-free core has reached 0.77$M_{\odot}$.
As discussed in \citet{s97} the chemical profile in the inner region of the He-exhausted core,
with constant abundances of C and O, is determined by the maximum extension
of the central He-burning convective region. Moving outwards from the centre, 
the peak in the oxygen abundance (and the corresponding depression of the C abundance) is 
produced when the He-burning shell crosses
the semiconvective region partially enriched in C and O, and $^{12}{\rm C}$ is converted into
$^{16}{\rm O}$ through the $^{12} {\rm C}(\alpha, \gamma) ^{16}{\rm O}$ reaction.
Beyond this region, the oxygen profile is built by 
the He-burning shell moving outwards. Gravitational contraction 
increases temperature and density in the shell, and given that
the ratio between the $^{12} {\rm C}(\alpha, \gamma) ^{16} {\rm O}$ and $3\alpha$ reaction
rates is lower for larger temperatures, the oxygen abundance
steadily decreases in the external part of the CO core.
The flat abundance profile of Ne is caused by the processing of N in the He core 
during He-burning through 
the reactions $^{14}\mathrm{N}\,( \alpha ,\gamma ) {}^{18}\mathrm{F}\,( \beta ^{+}) {}^{18}\mathrm{O}\,( \alpha ,\gamma ) {}^{22}\mathrm{Ne}\,$; as a result, the final Ne mass fraction is about equal to
the initial metal mass fraction $Z$ of the progenitor models.

The lower panel of Fig.~\ref{fig:bastiprofile} shows the chosen initial chemical stratification of
the corresponding WD, after the chemical rehomogeneization
of the inner core caused by the molecular weight inversion due to the above-mentioned local peak in the 
O abundance (and dip of the carbon mass fraction) discussed in \citet{s97}.

This kind of chemical profiles is admittedly \lq{artificial\rq}, because the CO-He-H chemical transitions
are not predicted to be sharp by fully evolutionary calculations \citep[see, e.g.][for some examples]{renedo, dge18};
This can be seen also in the upper panel of the same figure, which  displays the beginning of a slowly rising He
abundance from the boundary of the He-free core, with C and O abundances different from zero in a region where
He is also present.
In Sect.~\ref{test_profiles} we will show that imposing these sharp transitions instead of more realistic ones
does not have any major effect on the cooling timescales of the models \citep[but they have an impact on the model 
asteroseismic properties, see, e.g.,][]{gia}.

\begin{figure}
	\includegraphics[width=\columnwidth]{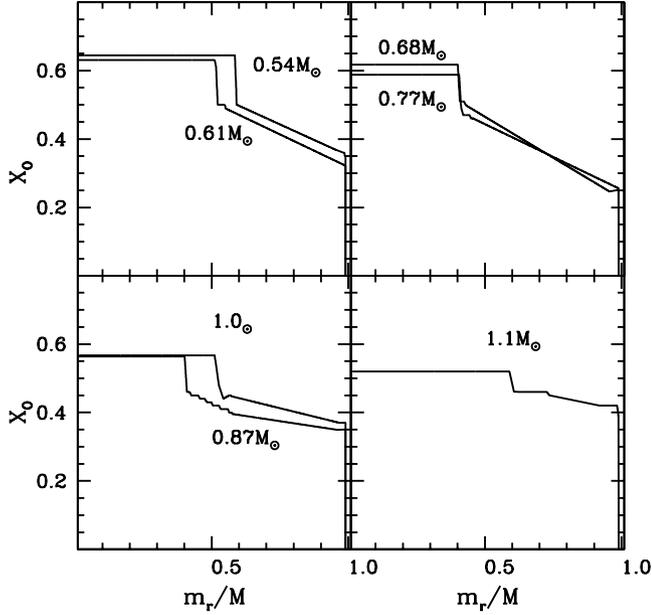}
        \caption{Oxygen abundance (mass fraction) stratification adopted for our models
          (see text for details).}
    \label{fig:bastiprofile_all}
\end{figure}

In this way we have determined the initial chemical profiles for all selected values of $M_{\rm WD}$, 
from calculations of progenitor models with $Z$=0.017.
Figure~\ref{fig:bastiprofile_all} displays the resulting oxygen stratifications (given in mass fraction $X_{\rm O}$)
across the CO cores, with the central value of $X_{\rm O}$  
typically decreasing with increasing $M_{\rm WD}$.
The corresponding carbon mass fraction $X_{\rm C}$ is given
by $X_{\rm C}$=1-$X_{\rm O}$-$Z$, where $Z$ is the progenitor metallicity, which is equal to a very good
approximation to the mass fraction of Ne ($X_{\rm Ne}$) across the core.

Given the importance of the amount of Ne in the core for the cooling timescales of
the models \citep[see, e.g.,][]{gb08}, for each WD mass we have calculated
models for five different initial Ne flat abundance profiles, corresponding to progenitors' 
metallicities equal to $Z=$0.006, 0.01, 0.017, 0.03, 0.04. 
When the mass fraction of Ne is below 0.006, we found that the effect on
both H-atmosphere and He-atmosphere models' cooling timescales is always negligible, hence 
we have also computed an additional set with no Ne,
representative of WDs from progenitors with $Z<$0.006.
For all these sets of calculations with varying Ne abundances,
we have kept the same CO profiles determined as described before (see Sect.~\ref{test_profiles}).

  Before closing this section, we wish to briefly comment on the results by \citet{giamcharp}, later refined by
  \citet{charp}, who have determined the radial
  chemical stratification of the $\sim$0.56$M_{\odot}$
  pulsating He-atmosphere WD KIC08626021, using asteroseismic techniques\footnote{See also the preliminary
  results by \citet{charproc} for five more objects.}. In the helium-free layers they
  found the presence of a central homogeneous core with a
  mass of $\sim$0.45 $M_{\odot}$ and $X_{\rm O} \sim$0.84, and also an external, very thin (in mass) 
  almost pure carbon buffer \citep[see Fig.~3 in][]{charp}. As discussed extensively by \citet{dge}, within our current
  understanding of the evolution of WD
  progenitors, it is difficult to reproduce the chemical structure of this star as derived from
  asteroseismology. For example, looking at the oxygen stratification of our 0.54$M_{\odot}$ models (a mass very close
  to the value estimated for KIC08626021) shown in Fig.~\ref{fig:bastiprofile_all},
  the central $X_{\rm O}$ is much lower than the asteroseismic value, and the chemically homogeneous inner core
  has a mass of $\sim$0.32$M_{\odot}$ instead of 0.45 $M_{\odot}$. We have assessed the impact of this
  asteroseismic CO 
  profile on our WD models by computing the evolution of 0.54$M_{\odot}$ WDs 
  with hydrogen and helium atmospheres
  respectively (the thickness of the H and He envelopes is the same adopted in the calculations
  described in this section) and no Ne  --using \citet{cas07}
  conductive opacities-- employing the CO stratification determined by \citet{charp}.
  We have found small differences in the cooling times compared to our calculations with the chemical 
  profile of Fig.~\ref{fig:bastiprofile_all}, amounting at most to 4\%
  (cooling times being shorter with the asteroseismic chemical stratification)
  at luminosities log($L/L_{\odot}$)$\sim-$4.9.

\subsection{The effect of Ne diffusion}
\label{neon} 

As discussed in detail in the literature \citep[see, e.g.][]{bh01,gb08},
$^{22}{\rm Ne}$ nuclei have a larger mass-to-charge ratio than $^{12}{\rm C}$ and $^{16}{\rm O }$ (the ratio between
atomic weight and charge is larger than 2 for $^{22}{\rm Ne}$); this results in a net downward
gravitational force on Ne, with a consequent slow diffusion towards the centre 
in the liquid layers, and the release of gravitational energy that contributes to the WD energy budget. 

\begin{figure}
	\includegraphics[width=\columnwidth]{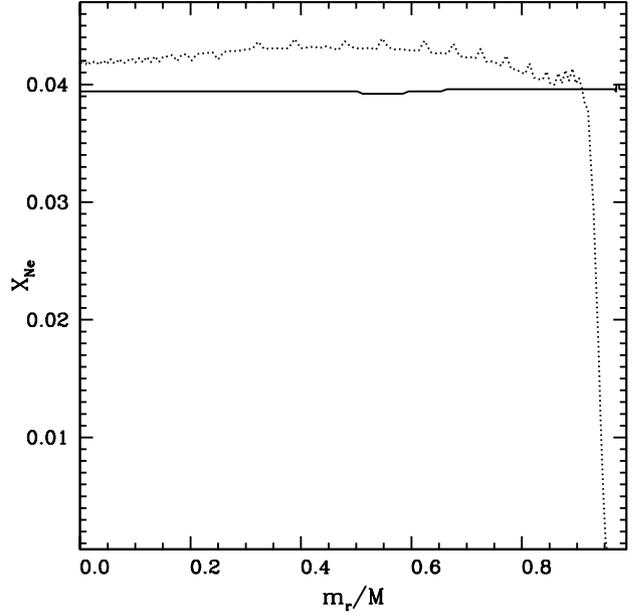}
        \caption{Initial (solid line) and final (dotted line) Ne abundance stratification
          (abundance in mass fraction) of a 0.61$M_{\odot}$ H-atmosphere model from a
          progenitor with $Z$=0.04 
          (see text for details).}
    \label{fig:neprofile}
\end{figure}

Figure~\ref{fig:neprofile} displays the initial Ne abundance profile in the CO core of our
0.61$M_{\odot}$ H-atmosphere calculations
with a $Z$=0.04 progenitor, and the final Ne stratification when the core has fully crystallized.
On the whole, the final distribution of Ne is more concentrated towards the centre
compared to the initial one,
and the details of the final abundance profile arise from the interplay between the settling 
of Ne towards the centre,
and the inhibition of diffusion at the crystallization front, that moves outwards with time.

\begin{figure}
	\includegraphics[width=\columnwidth]{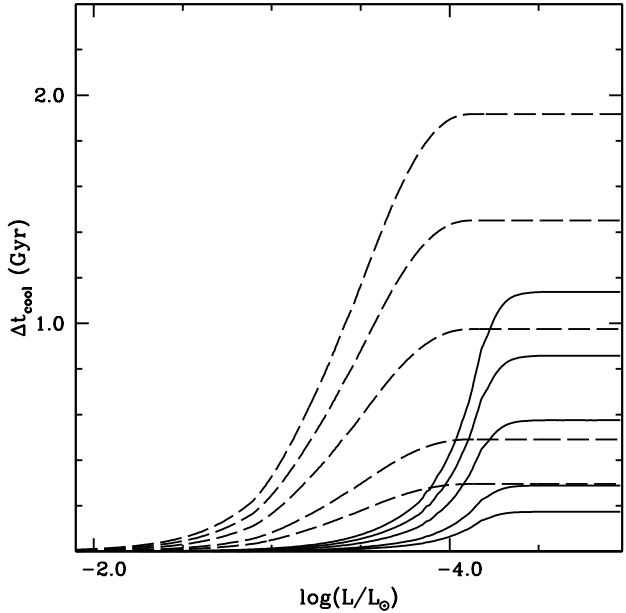}
        \caption{Cooling time delay $\Delta t_{\rm cool}$ (in Gyr) as a function of the 
          luminosity, caused by the diffusion of Ne in 
          0.61$M_{\odot}$ (solid lines) and 1.0$M_{\odot}$ (dashed lines) H-atmosphere WD models, 
          from progenitors with $Z$=0.006, 0.01, 0.017, 0.03, 0.04 (in order of increasing
          maximum $\Delta t_{\rm cool}$) respectively.}
    \label{fig:nedelayDA}
\end{figure}

\begin{figure}
	\includegraphics[width=\columnwidth]{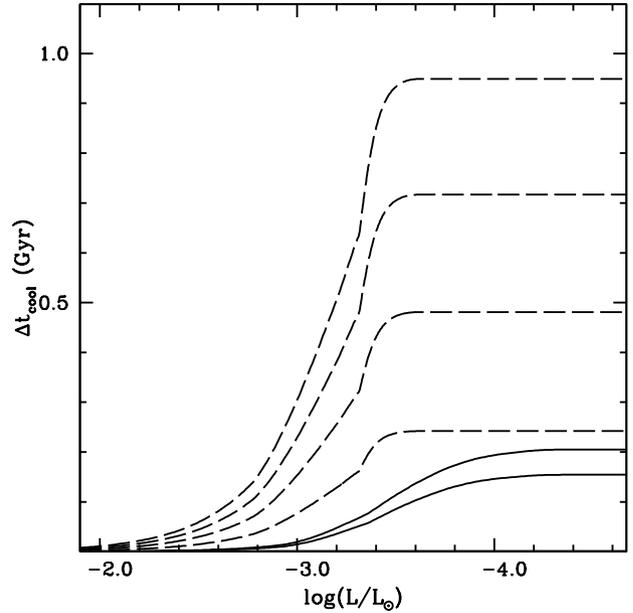}
        \caption{As Fig.~\ref{fig:nedelayDA} but for the corresponding He-atmosphere models.
          The displayed 0.61$M_{\odot}$ calculations are for progenitors with $Z$=0.03 and 0.04; the
          1.0$M_{\odot}$ calculations are for progenitors with $Z$=0.01, 0.017, 0.03, 0.04 (in order of increasing
          maximum $\Delta t_{\rm cool}$).
          The cooling delay for models from lower metallicity progenitors are negligible.}
    \label{fig:nedelayDB}
\end{figure}

As already mentioned, this displacement of Ne towards the centre causes a release of gravitational energy, that slows
down the cooling of the models. This is shown by Figures~\ref{fig:nedelayDA} and \ref{fig:nedelayDB},
which display the extra time
$\Delta t_{\rm cool}$ that selected H-atmosphere and He-atmosphere models --calculated with \citet{cas07} electron
conduction opacities-- take to cool down to a given luminosity, due to the diffusion of Ne.
In addition to the mass of the model,
the progenitor metallicity here plays a major role, because an increase of the amount of Ne in the core increases
$\Delta t_{\rm cool}$. These delays can have the same order of magnitude of the effect of 
latent heat release plus phase separation upon crystallization of the CO mixture (which amounts
to $\sim$1.85~Gyr for the 0.61$M_{\odot}$ H-atmosphere models), depending on the initial Ne abundance.

The values of $\Delta t_{\rm cool}$ due to Ne diffusion (and also crystallization) are in general smaller
for He-atmosphere models,
due to the lower opacities of the non-degenerate envelope, and for the lower mass models they are already negligible
for progenitors around solar metallicity.

\section{Tests on the initial chemical profiles}
\label{test_profiles}

In the previous section we have compared the cooling times of WD models from progenitors with different
initial metallicities. These models have been computed with Ne abundances appropriate for the progenitor metallicity,
but with a fixed CO stratification at a given $M_{\rm WD}$, determined from progenitors with metallicity
about solar and mass derived from a single IFMR.
In this way we have quantified the effect of Ne diffusion on the cooling times of our models,
for different Ne abundances and $M_{\rm WD}$.

In this section we want first to assess whether changing just the Ne abundance in the core is a reasonable approximation
to determine reliable cooling times for WDs produced by progenitors with varying metallicity.
Whilst the Ne mass fraction is always to a very good approximation
equal to $Z$ and uniform across the core for any $M_{\rm WD}$, irrespective of the assumed values of $M_{\rm i}$,
the situation for CO is more complex.
In fact, the CO stratification for a given $M_{\rm WD}$
depends in principle on the progenitor $Z$ through a combination 
of the effect of $Z$ on the CO profiles at fixed $M_{\rm i}$, and the effect of metallicity on the IFMR, that  
associates a progenitor mass $M_{\rm i}$ to a given $M_{\rm WD}$.

Semiempirical determinations of the IFMR are based on objects of  
metallicity typically around solar, apart from the lowest mass WDs included in \citet{cummings} IFMR adopted here,
that come from observations of the globular cluster NGC~6121.
Therefore, we do not have empirical indications of how the IFMR globally changes with changing progenitor metallicity,
with the consequent not well established impact on the CO stratification of the WD models.

As already mentioned, theoretical predictions of the IFMR do not provide consistent results. The only qualitative
trend on which various theoretical calculations of asymptotic giant branch evolution \citep[see, e.g.,][]{mgir, wf}
generally agree is
that, with decreasing $Z$, the value of $M_{\rm i}$ for a fixed $M_{\rm WD}$ tends typically to decrease, at least for
$M_{\rm i}$ above $\sim$2-2.5$M_{\odot}$. 
Assuming this qualitative trend is real, we have performed the following test on H-atmosphere models, calculated with
the \citet{cas07} conductive opacities.

\begin{figure}
	\includegraphics[width=\columnwidth]{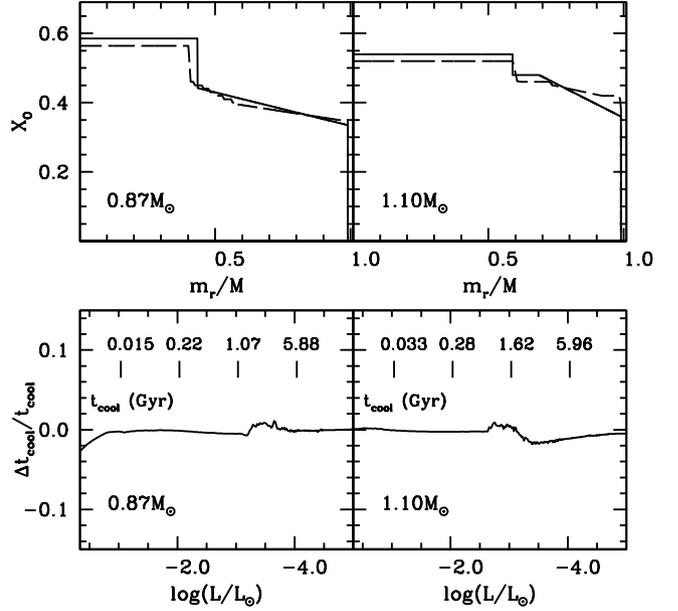}
        \caption{{\sl Upper panels:} Pairs of oxygen abundance 
          profiles (abundance in mass fraction) employed for the calculation of
          0.87 and 1.1$M_{\odot}$ H-atmosphere WD models. The dashed
          lines display our adopted reference profiles, whilst the solid lines correspond to
          the case of metal poor progenitors with $Z$=0.0002 (see text for details).
          {\sl Lower panels:} Relative difference between the cooling times obtained using the CO stratification of the
          metal poor
          progenitor, and those obtained from the reference stratification, as a 
          function of the luminosity. Selected cooling ages from the calculations with the reference profiles are marked.}
    \label{fig:testprofiles}
\end{figure}

The upper panels of Fig.~\ref{fig:testprofiles} display pairs of oxygen abundance profiles employed to
calculate 0.87$M_{\odot}$ and 1.1$M_{\odot}$ models, respectively.
We have considered first our reference abundances determined as described in the previous section, which in this case
come from progenitors with $M_{\rm i}$=3.6$M_{\odot}$ (for $M_{\rm WD}$=0.87$M_{\odot}$), and
$M_{\rm i}$=6.4$M_{\odot}$ (for $M_{\rm WD}$=1.1$M_{\odot}$), both calculated with $Z$=0.017.
For these progenitor the mass of the He-free core
has increased by, respectively, $\sim$0.16$M_{\odot}$ and $\sim$0.18$M_{\odot}$ compared to the values at the first thermal pulse.

The second set of oxygen stratifications have been obtained by considering progenitors with $Z$=0.0002 
(a low metallicity typical of metal-poor
Galactic globular clusters) and masses $M_{\rm i}$ such that the models reach $M_{\rm WD}$=0.87$M_{\odot}$
and 1.1$M_{\odot}$ when the mass of their He-free cores had grown 
by $\sim$0.16$M_{\odot}$ and $\sim$0.18$M_{\odot}$, respectively, compared to the values at the first thermal pulse,
like the case of our reference IFMR.
The corresponding values of $M_{\rm i}$ are equal to 2.5$M_{\odot}$ and 5$M_{\odot}$, lower than the values prescribed by our 
reference IFMR, consistent with the qualitative trend expected from
asymptotic giant branch evolutionary models.

For each value of $M_{\rm WD}$, the two abundance profiles from progenitors with different metallicities turn out to be
very similar.
In the assumption that our assumed metal poor IFMR is the \lq{real\rq} one, then 
we can expect that if we use our reference CO stratification to calculate models for WDs in $Z$=0.0002
populations, we will determine cooling times negligibly
different from the case of using the appropriate metal poor IFMR.

This is shown in the lower panels of the same figure, which compare the cooling times of our reference
0.87$M_{\odot}$ and 1.1$M_{\odot}$ 
calculations without Ne (because its effect on the cooling times is negligible at this metallicity) with models 
calculated using the core stratification from the appropriate $Z$=0.0002 progenitors
(and no Ne).
We find almost negligible fractional differences of the cooling times, the largest values being at most equal to 2\%.

Given the results of this test --based on necessarily arbitrary assumptions-- 
that somehow mimics the qualitative behaviour of the IFMR with $Z$ as predicted by
theoretical asymptotic giant branch models, and the lack of empirical assessments of how the IFMR
varies with metallicity, we consider our models calculated with fixed IFMR and CO profiles and varying Ne abundances 
(see Sect.~\ref{mod}) to be an acceptable choice 
to study the cooling of WDs in populations with different metallicities.

\subsection{Testing the idealized chemical interfaces}
\label{transitions}

Another assumption --common in the literature-- made in our WD model calculations is that the CO-He-H chemical transitions
are sharp, with an abrupt switch from the CO (and Ne) core to a pure-He envelope, and from the He envelope to a pure-H envelope. 
As mentioned in the previous section, this is just an idealization. In reality, progenitor model calculations show
a narrow (in mass) transition region around the He-free core, where
the He abundance progressively increases outwards, and the C and O abundances drop to zero, due to incomplete burning
in the He-shell.
Additionally, H and He are not \lq{separated\rq} in the envelopes of the asymptotic giant branch progenitors,
even considering the effect of atomic diffusion during the previous evolution, and pure-H and He envelopes
do not account for the metals in the initial chemical composition of the progenitor. The CNO of the original chemical composition
can be involved in hydrogen burning if the mass of the hydrogen layers is large enough \citep[see, e.g.,][and below]{a15}. 

As discussed for example in \citet{renedo}, atomic diffusion during the WD cooling makes helium and metals sink
from the outer evelope, producing a thin (in mass) pure-H layer that thickens as evolution proceeds.
Chemical transitions exhibit increasingly less sharp discontinuities, and the metals progressively sink 
towards deeper and deeper regions of the envelope.
Overall, the chemical stratification of WD model envelopes is expected not only to be initially without sharp chemical
transitions,  
but also to change slowly during the cooling. 
To test the effect of the simplified chemical transitions adopted in our models on the predicted cooling ages, 
we have performed the following numerical experiment. 

\begin{figure}
	\includegraphics[width=\columnwidth]{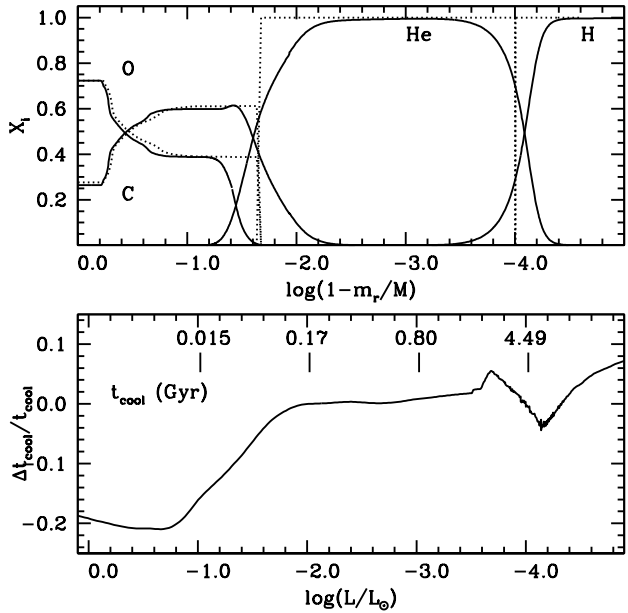}
        \caption{{\sl Upper panel:} Solid lines display the C, O, He and H stratification
          (abundances in mass fraction) 
          at the beginning of the cooling sequence of 0.61$M_{\odot}$ H-atmosphere 
          models by \citet{renedo}. The dotted lines display our adopted core stratification with CO profile determined as
          described in Sect.~\ref{chemistry}, plus stratified pure-H and He envelopes which match
          the total mass of H and He 
          of \citet{renedo} models. {\sl Lower panel:} Relative difference
          of cooling times as a function of luminosity with respect to \citet{renedo} calculations, 
          of 0.61$M_{\odot}$ models calculated using the
          stratification with pure-He and H envelopes shown in the upper panel   
          (dotted line -- see text for details). Some reference cooling ages from \citet{renedo} calculations are marked.}
    \label{fig:lpcode}
\end{figure}

We have considered the C, O, He, H chemical stratification of the 0.61$M_{\odot}$ H-atmosphere WD model
by \citet{renedo}, calculated
from a progenitor with $M_{\rm i}$=2$M_{\odot}$ and metallicity $Z$=0.01, taken at log($L/L_{\odot}$)$\sim$1.0, and displayed
in the upper panel of Fig.~\ref{fig:lpcode}. In this plot the mass fractions of C and O below $10^{-4}$ found in 
the He- and H-rich envelopes of \citet{renedo} model are set to zero; the
nitrogen abundance, that is accounted for by \citet{renedo} is also set to zero everywhere.
The total masses of hydrogen and helium in the envelope
correspond to $q({\rm H})$=$1.3 \times 10^{-4}$ and $q({\rm He})$=$2.3 \times 10^{-2}$, respectively.

Starting from this chemical profiles, we have built the second chemical stratification displayed in 
the upper panel of Fig.~\ref{fig:lpcode}, which mimics the way we have built the chemical make-up of our new BaSTI 
models presented in this paper (see Sect.~\ref{chemistry}).
The CO-He-H discontinuities are sharp, the envelope is made
of pure-H and pure-He layers with $q({\rm H})$=$1.3 \times 10^{-4}$ and $q({\rm He})$=$2.3 \times 10^{-2}$, to preserve 
the total amounts of H and He of  
\citet{renedo} initial model, and the CO stratification has been obtained by taking the CO abundance profile inside
the He-free core of \citet{renedo} model. Given that 
we do not account for the small amount of C and O in the abundance tails that overlap 
the helium envelope (i.e., in the region between log($1-m_r/M$) between $\sim -$1.2 and $\sim -2.4$) we needed
to apply a very small rescaling of the mass coordinate in the CO core, to conserve the total mass
of the model. This profile is a proxy for \citet{renedo} initial profile, built essentially like
the chemical stratification of the set of new BaSTI WD models we are presenting in this paper.

Using this chemical stratification, we have calculated with our
code the evolution of a 0.61$M_{\odot}$ WD model using the same EOS, boundary conditions, radiative and conductive
opacities, neutrino energy loss rates, and CO phase diagram employed by \citet{renedo}, who used the LPCODE for their 
calculation.
For this test --like for all our BaSTI WD models-- we neglect the evolution of the H and He profiles
due to atomic diffusion, set all metal abundances to zero in the non-degenerate envelope, and hence
we neglect also the occurrence --a small effect in \citet{renedo} calculations-- of CNO burning in the hydrogen
envelope.

The lower panel of Fig.~\ref{fig:lpcode} displays the relative difference of the cooling times
(zeroed at log($L/L_{\odot}$)$\sim$1.0) with respect to \citet{renedo} models\footnote{The result of this
  comparison is significant, given that 
in \citet{salaris13} we have shown how WD models calculated with the LPCODE and our BaSTI code using 
the same physics inputs and initial chemical stratification,
have cooling times within 2\% or better, and radii consistent at the 0.5\% level.}.
Our calculation displays systematically shorter ages  
(up to about 20\% shorter) during the early phases of cooling, when the age of the models is on the order of 10~Myr
or less. This difference is most likely due to the different thermal structures 
at log($L/L_{\odot}$)$\sim$1.0, the start
of our calculations\footnote{\citet{renedo} followed the full evolution from the main sequence
  to the white dwarf stage of the progenitor, whilst we have converged an initial \lq{artificial\rq}
  structure at log($L/L_{\odot}$)$\sim$1.0, with the chosen total mass and chemical profile.}.
By log($L/L_{\odot}$)$\sim -$1.8 and cooling ages around 100~Myr neutrino emission --by now inefficient-- has erased
any difference of the initial thermal structure of the models, and the cooling times become almost identical.
Differences reach at
most about $\pm$6\% when log($L/L_{\odot}$) goes below $\sim -$4.0, corresponding to ages above $\sim$4.5~Gyr.
The radius of our models is systematically lower than \citet{renedo} results, differences amounting to just about 3\%
at log($L/L_{\odot}$)=0.5, decreasing to 1.5\% and log($L/L_{\odot}$)=$-$2.0, and below 1\% when  log($L/L_{\odot}$)
is lower than $\sim -3$.

To summarize, this test shows that the use of sharp chemical interfaces, the neglect of the 
abundance of the metals in the
H- and He-rich layers, and the effect of atomic diffusion and CNO burning in the envelope of models with
total H and He mass fractions similar to the choice adopted in our BaSTI calculations, have a minor effect on
the models' cooling times and radius, amounting to at most a few percent.

\section{The role of the electron conduction opacities}
\label{compopa}

\begin{figure}
	\includegraphics[width=\columnwidth]{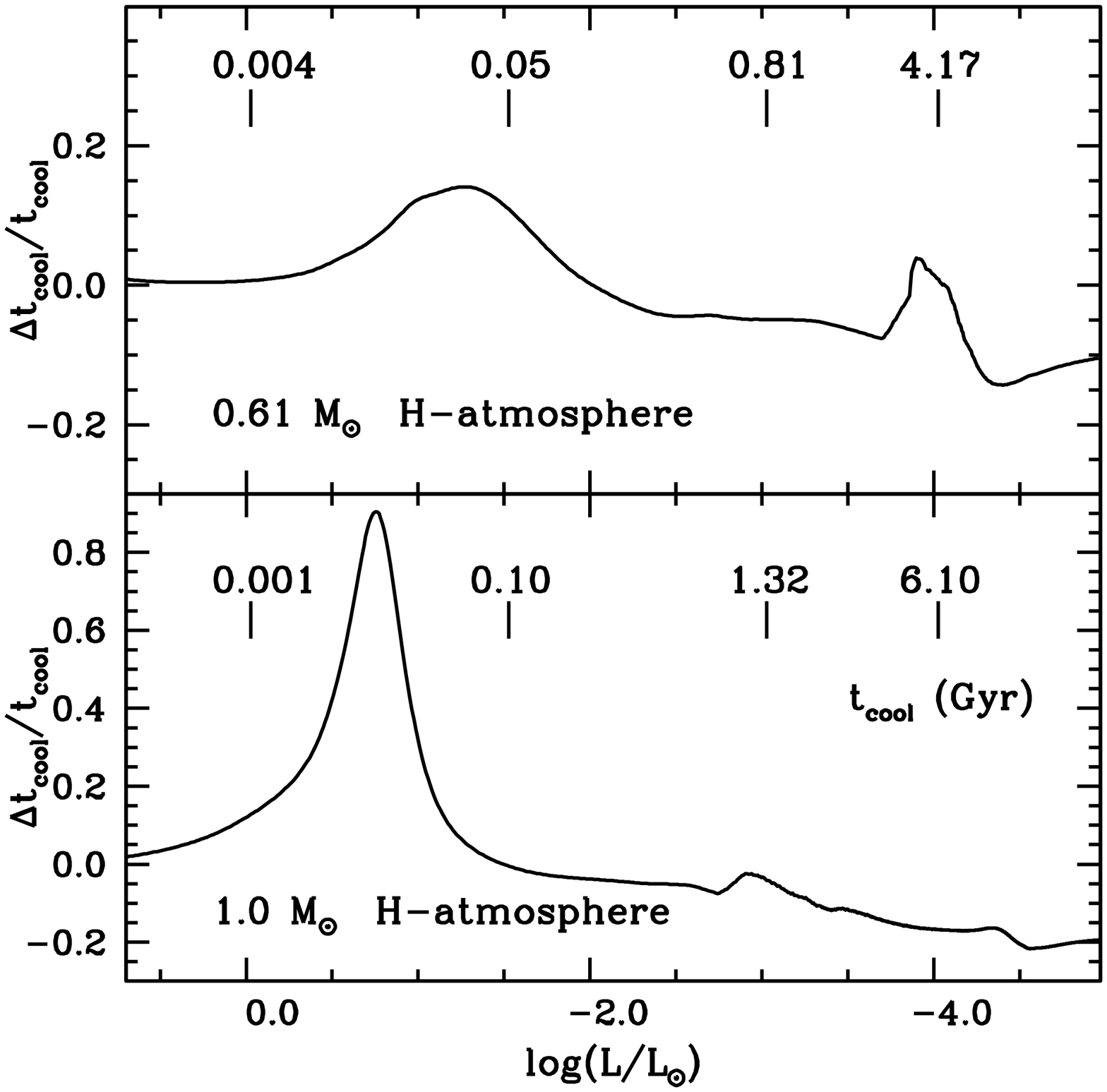}
        \caption{Relative difference
          of cooling times as a function of luminosity for 
          our 0.61$M_{\odot}$ and 1.0$M_{\odot}$ H-atmosphere models with no Ne (corresponding to metal poor populations), 
          calculated with \citet{b20} and \citet{cas07} opacities, respectively. A positive difference means that
          models computed with \citet{b20} opacities have longer cooling times.
          Some reference cooling ages from the calculations with \citet{cas07} opacities are marked.}
    \label{fig:opada}
\end{figure}

\begin{figure}
	\includegraphics[width=\columnwidth]{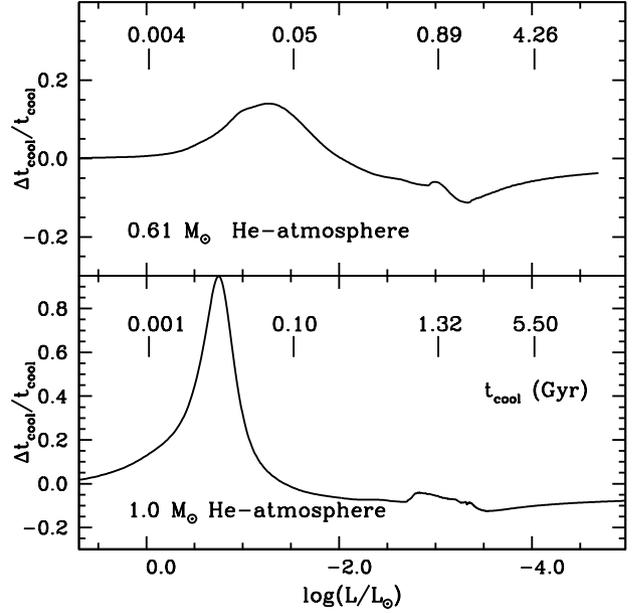}
        \caption{Same as Fig.~\ref{fig:opada} but for He-atmosphere models.}
    \label{fig:opadb}
\end{figure}

As motivated and discussed in Sect.~\ref{physics}, we have computed two sets of new BaSTI WD calculations, employing 
the \citet{cas07} and \citet{b20} conductive opacities, respectively.
\citet{b20} and \citet{cpsp} have shown that WD models calculated with \citet{b20}
conductive opacities display longer cooling times at bright luminosities and shorter cooling times at faint luminosities,
compared to calculations with \citet{cas07} opacities;  Figs~\ref{fig:opada} and \ref{fig:opadb} confirm their
results, displaying the relative difference of the 
cooling ages of 0.61$M_{\odot}$ and 1.0$M_{\odot}$ H-atmosphere and He-atmosphere models calculated with both sets of opacities.

The lower \citet{b20} conductive opacities for the H and He layers 
induce a faster cooling of the core, which in turns reduces the efficiency
of neutrino cooling at bright luminosities, and increases the cooling times at a fixed value of log($L/L_{\odot}$).
The absolute values of the cooling times at these luminosities are however 
fairly short, well below 100~Myr.
With decreasing luminosities, the cooling times calculated with \citet{b20} 
opacities become increasingly shorter than calculations with \citet{cas07} opacities, because of the faster
cooling of the structure, and differences increase with increasing $M_{\rm WD}$.
This trend is temporarily broken by the earlier start of CO crystallization in
the models with \citet{b20} opacities, and the associated earlier onset 
of the release of latent heat and extra energy due to phase separation.

For initial Ne abundances that have an impact on the cooling times of the models, the trends shown in
Figs.~\ref{fig:opada} and \ref{fig:opadb} are unchanged. Due to the lower opacity of the denser layers of the envelope, the
cooling delay caused by the diffusion of Ne in the liquid phase 
is shorter in models with \citet{b20} opacities, but the increase of the relative age  
differences at low luminosities is however small, by at most a few percent for the higher masses and higher Ne
abundances.

\section{Sedimentation of iron in the core}
\label{iron}

After C, O and Ne, the next more abundant element in the CO core is predicted to be $^{56} {\rm Fe}$ (hereafter Fe), part of
the initial metal content of the WD progenitors.
For the solar scaled metal mixture by \citet{caffau} used in our \citet{bastiiacss} progenitor calculations,
the mass fraction of Fe is equal to about 8.8\% of $Z$.

\citet{caplanFe} have very recently computed new phase diagrams for a COFe ternary mixture 
and performed molecular dynamics calculations, showing the Fe present in WD CO cores can precipitate to form 
an inner iron core during crystallization. This process would release energy (like the diffusion of Ne,
or the phase separation of the CO mixture) slowing down the cooling of the models.
The sedimentation of Fe was already briefly explored by \citet{sc93} and 
\citet{segretain94}, and discussed in more details by 
\citet{xu}, making use of calculations of earlier phase diagrams.
The details of this process are still uncertain, and various outcomes are possible regarding the chemical stratification of
the core during the crystallization of the C, O and Fe mixture \citep[see][]{caplanFe}; here we discuss the 
more {\sl standard} scenario presented by \citet{caplanFe}, making a very
preliminary estimate of the effect of Fe sedimentation on the cooling times of our models.

When $\Gamma_{\rm C}=6^{5/3} (2.275 \, 10^5/T) (\rho/2)^{1/3}$ (with density given in g/$\rm cm^3$ and temperature in K --
the Coulomb parameter of a pure-C composition) reaches a critical value equal to 140 in a layer
of the CO core, its Fe precipitates on very short timescales towards the centre; we assume here that the fraction
of Fe that precipitates is 100\% \citep[it could be less than this, according to][]{caplanFe}
and that this iron settles
virtually instantaneously at the centre,
replacing C and O \citep[and we assume also Ne, although the fate of Ne is not discussed in the
  paper by][and another possibility is that
  this core containing the precipitated Fe is made instead of a C, O, Fe alloy]{caplanFe}.
As the cooling proceeds, more external layers
reach $\Gamma_{\rm C}$=140, and the mass of this pure-Fe core increases, until the CO mixture starts to crystallize
and Fe sedimentation stops \citep[][speculate that Fe sedimentation might continue also after C and O start to
crystallize]{caplanFe}.
From this moment on, the CO crystallization proceeds like in the case of neglecting the presence of Fe.

We have implemented this process to follow the evolution of our  
$M_{\rm WD}$=0.54$M_{\odot}$ and $M_{\rm WD}$=1.0$M_{\odot}$ models, respectively, coming from progenitors with $Z$=0.04, to maximize
the amount of Fe in the core.  This metallicity  
corresponds to a Fe mass fraction $X_{\rm Fe}$=0.0035. In the calculations we assumed that the diffusion of Ne continues
also during the sedimentation of Fe.

For the $M_{\rm WD}$=0.54$M_{\odot}$ case 
all iron within the inner 11\% in mass of the CO core is displaced to form a pure-Fe 
core with mass $M_{\rm Fe}$=0.00021$M_{\odot}$, before the CO component starts to crystallize (this happens
when $\Gamma_{\rm C} \sim 160$ in the innermost CO layers outside the Fe core).
The layers depleted in Fe are within the 
flat part of the CO abundance stratification, that spans
the inner 55-60\% in mass of the core (see Fig.~\ref{fig:bastiprofile}).

The C, O and Ne removed from the layers now belonging to the iron core replace the heavier
Fe in the layers affected by sedimentation, and this causes
a small inversion of the mean molecular weight with the overlying Fe-rich layers.
The whole region beyond the iron core that originally had an homogeneous CO stratification gets 
mixed by convection to cancel this inversion, and part of the energy gained by the sedimentation of Fe
is absorbed in these convective layers\footnote{Also during the CO crystallization part of the 
  energy gained is absorbed by the layers that are rehomogeneized by
  the convective episodes caused by phase separation \citep[see,e.g.,][]{isern00}.}.
As a result, the C and O abundances in these mixed
layers are only very slightly changed, by much less than 1\%. The Ne abundance profile, which is affected by diffusion, 
is rehomogeneised in these layers, but the effect on the contribution of the further development of Ne diffusion 
to the model energy budget is negligible.

The cooling age of the models increases by $\sim$150~Myr in the range between
log($L/L_{\odot}$) $\sim -$3.8 and $-$3.9 when we use the \citet{b20} opacities (at cooling ages around 2.5~Gyr),
and by about 190~Myr between  log($L/L_{\odot}$) $\sim -$3.9 and $-$4.0 in models
calculated with \citet{cas07} opacities (at cooling ages around 3.5~Gyr).

For the $M_{\rm WD}$=1.0 calculations the Fe core mass at the end of the sedimentation is equal to 
$M_{\rm Fe}$=0.00057$M_{\odot}$, equivalent to about 16\% of the total Fe in the CO core; as a result, the cooling age
increases by $\sim$120~Myr in the range between  log($L/L_{\odot}$) $\sim -$2.6 and $-$2.8 in models
calculated with \citet{b20} opacities (at cooling ages around 790~Myr), and by about 200~Myr between log($L/L_{\odot}$) $\sim -$2.7 
and $-$2.9 in models calculated with \citet{cas07} opacities (at cooling ages around 1~Gyr).

\section{Comparison with other models}
\label{other}

\begin{figure}
	\includegraphics[width=\columnwidth]{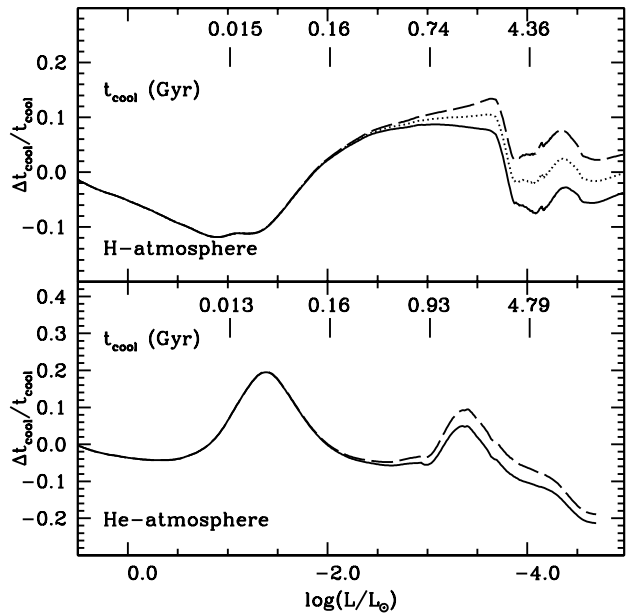}
        \caption{Relative difference of cooling times (in Gyr) as a function of luminosity
          with respect to the 0.61$M_{\odot}$ H-atmosphere calculations from S10 of, respectively,
          our new 0.61$M_{\odot}$ H-atmosphere 
          calculations with \citet{cas07} opacities without Ne diffusion (solid line), with Ne diffusion and
          $Z$=0.017 Ne abundance (dotted line), with Ne diffusion and
          $Z$=0.04 Ne abundance (dashed line). Some reference ages from S10 calculations are marked.}
    \label{fig:oldnewa}
\end{figure}

\begin{figure}
	\includegraphics[width=\columnwidth]{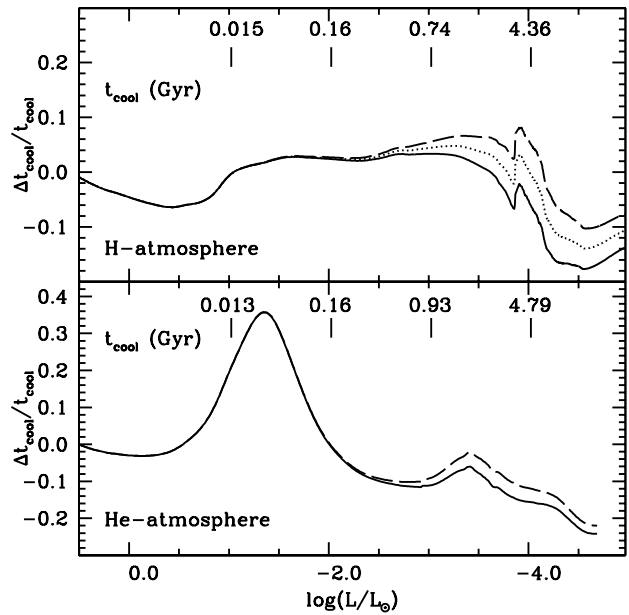}
        \caption{As Fig.~\ref{fig:oldnewa}, but for our new models calculated with
          \citet{b20} opacities.}
    \label{fig:oldnewb}
\end{figure}

To put our new calculations in the broader context of existing WD models, we make first a
comparison with our previous S10 results. While the models' radii change by less than 1\% compared to S10,
the same is not true regarding the cooling times. 
As an example, Figs.~\ref{fig:oldnewa} and \ref{fig:oldnewb} compare the cooling times
of our 0.61$M_{\odot}$ H-atmosphere and He-atmosphere calculations with their S10 counterparts,
which do not include Ne diffusion.

For log($L/L_{\odot}$) larger than about $\sim -$ 3.5, the $\pm 10$\% differences of cooling times with our models
without Ne diffusion 
are due just to the older conductive opacities adopted by S10. Below
this luminosity we see the effect also of the new CO phase diagram employed here. The \citet{bd21} diagram
delays the onset of crystallization
compared to the \citet{sc93} diagram used by S10, and the amount of matter redistributed by phase separation
for a given $M_{\rm WD}$ is sizably smaller when the \citet{bd21} phase diagram is employed.
This causes a smaller amount of energy released by CO phase separation,
causing a faster cooling compared to S10 models\footnote{The differences
  with S10 results due to the use of \citet{bd21} phase diagram instead of the \citet{sc93} one are very similar
  to the results discussed by \citet{althaus12}. These authors compared models calculated with \citet{sc93} phase diagram and
  the \citet{horowitz} one, which is very similar to the more recent \citet{bd21} diagram.}.
The CO stratification
is also not the same as in S10, although differences are small, and the impact on the crystallization process 
is relatively minor.
When we include Ne diffusion the cooling times increase and eventually, in models calculated with \citet{cas07}
opacities and for super-solar Ne abundances, they become longer than S10 results
when log($L/L_{\odot}$) is below $\sim -$4. In case of
the He-atmosphere models, for both choices of conductive opacities --even when including Ne diffusion-- the cooling times
of our models at log($L/L_{\odot}$) below $\sim -$4 are always shorter than S10 results. 
 
\begin{figure}
	\includegraphics[width=\columnwidth]{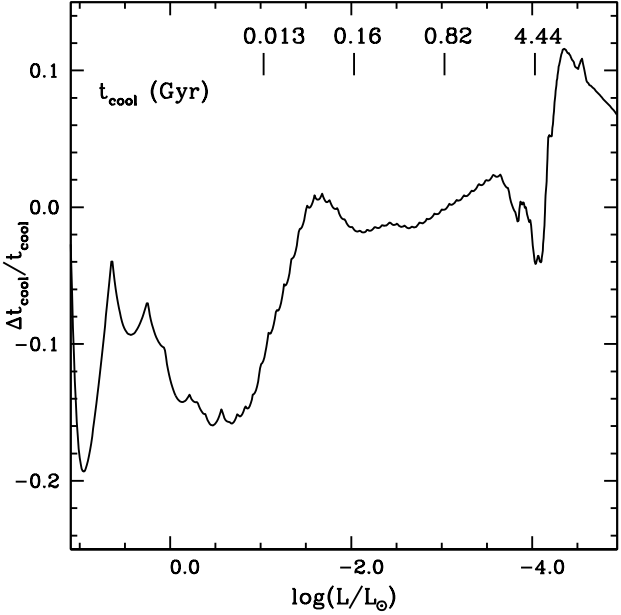}
        \caption{Relative difference of the cooling times (zeroed at log($L/L_{\odot}$)=1.1) as a function of luminosity
          between our 0.61$M_{\odot}$ H-atmosphere cooling track
          with Ne abundance from a $Z$=0.017 progenitor, and a 0.6$M_{\odot}$ H-atmosphere track with solar
          metallicity progenitor from
          \citet{camisassa16}. Some reference cooling ages from \citet{camisassa16} calculations are marked.}
    \label{fig:camisassa}
\end{figure}

In Fig.~\ref{fig:camisassa} we compare the cooling times of our 0.61$M_{\odot}$ H-atmosphere models
including Ne diffusion and a progenitor
metallicity $Z$=0.017, calculated with \citet{cas07} opacities, with results by 
\citet{camisassa16} for a 0.6$M_{\odot}$ H-atmosphere WD computed with the same conductive opacities,
which include Ne diffusion and account for the full evolution
of a solar metallicity progenitor\footnote{At \url{http://evolgroup.fcaglp.unlp.edu.ar/TRACKS/newtables.html}}.
\citet{camisassa16} models have a different thickness of the envelope layers
compared to our calculations, a different CO stratification
and there are some differences in the physics inputs, but we take the results of the comparison
of cooling times at face value.

At luminosities above log($L/L_{\odot}$) $\sim -1.5$ and ages below $\sim 100$~Myr our models display 
systematically lower cooling ages, most likely due to differences in the thermal structure 
at the beginning of the cooling evolution. Cooling times are then
within $\pm$2-4\% down to log($L/L_{\odot}$) $\sim -$4, beyond which our models evolve  
at a slower speed and display cooling times up to about 10\% longer.


\section{Summary}
\label{summary}

We have extended our updated BaSTI stellar evolution archive with the inclusion of new models for H- and
He-atmosphere CO-core WDs, computed using the C,O, and Ne stratification obtained from new progenitor calculations,
adopting a semiempirical IFMR.
We have also updated the model physics inputs compared
to the previous S10 BaSTI WD models; in particular, we now include Ne diffusion in the core, an updated CO 
phase diagram, and updated electron conduction opacities.

We have calculated the evolution of models with masses $M_{\rm WD}$ equal to 
0.54, 0.61, 0.68, 0.77, 0.87, 1.0 and 1.1 $M_{\odot}$, made of a CO core surrounded by pure-He layers of mass fraction 
$q({\rm H})=10^{-2}$, and (for H-atmosphere models) outermost
H layers enclosing a mass fraction $q({\rm H})$=$10^{-4}$.
For any given $M_{\rm WD}$ we have computed the evolution of models with several Ne abundances,
appropriate for initial metallicities
of the progenitors equal to $Z$=0.006, 0.01, 0.017, 0.03, and 0.04, corresponding to
[Fe/H]=$-$0.40, $-$0.20, 0.06, 0.30 and 0.45. We have also models calculated with no Ne, representative of
WDs from progenitors with [Fe/H] lower than $-$0.40 ($Z$=0.006), because the corresponding
small quantities of Ne have a negligible effect on the cooling times of the models.
In all these calculations we have kept fixed the CO abundance profiles to
those obtained from progenitors with $Z$=0.017 ([Fe/H]=0.06),
and the IFMR is kept also unchanged in all calculations, for it is not yet established empirically if/how the IFMR varies
as a function of the progenitor metallicity.
We estimated that the effect of this assumption on the model cooling times is potentially negligible, when we 
take into account the qualitative trend with metallicity of the IFMRs predicted by theoretical asymptotic giant branch models.

All our models comprise a CO (and Ne) core surrounded by pure-He and H (in H-atmosphere models) envelopes.
These sharp chemical transitions
and envelopes made of pure-He and H are actually a simplified description of the chemical structure of a WD.
We have performed test calculations which have quantified to at most $\pm$6\%
the effect on the model cooling times of our idealization of the chemical
transitions, and the neglect of the metals in the H and He envelopes.

We have also made a first, preliminary estimate of the effect of Fe sedimentation on the cooling times
of WD models, following the recent results by 
\citet{caplanFe} regarding the phase diagram of C, O and Fe mixtures. The effect turns out to be generally minor,
but there are still substantial uncertainties on the details of this process.

Two complete sets of calculations have been performed, for two different choices of the electron conduction opacities,
to reflect the current uncertainty in the evaluation of the electron thermal conductivity in the crucial
(for the H and He envelopes) transition regime between moderate and strong degeneracy.
Models calculated with different opacities indeed display non-negligible differences in their cooling times, larger for
higher values of $M_{\rm WD}$. In a next paper we will employ both sets of calculations to study WDs in old stellar populations,
with the aim of setting some stringent constraints on how to bridge calculations of conductive opacities
for moderate and strong degeneracy, as suggested by \citet{cpsp}.

We make publicly available the cooling tracks from both sets of calculations, including cooling times and magnitudes in
the Johnson-Cousins, Sloan, Pan-STARSS, Galex, $Gaia$-DR2, $Gaia$-eDR3, $HST$-ACS, $HST$-WFC3, and $JWST$
photometric systems.

\section*{Acknowledgements}
We thank the referee for constructive comments that helped us improving the manuscript.
SC acknowledges support from Premiale INAF MITiC, from
INFN (Iniziativa specifica TAsP), and from PLATO ASI-INAF agreement
n.2015-019-R.1-2018. 
MS acknowledges support from STFC Consolidated Grant ST/V00087X/1.
SC warmly thanks the Instituto de Astrofisica de Canarias for the hospitality and
the \lq{Programa de investigadores visitantes}\rq  de la Fundacion Jesus Serra.

\section*{Data Availability}

All WD tracks in several photometric systems are available at the new 
official BaSTI website \url{http://basti-iac.oa-abruzzo.inaf.it}.



\bibliographystyle{mnras}
\bibliography{WDpaper} 








\bsp	
\label{lastpage}
\end{document}